\documentclass[12pt]{article} 

\usepackage[utf8]{inputenc} 

\usepackage{geometry} 
\usepackage{float}
\usepackage[caption = false]{subfig}
\usepackage{graphicx}

\graphicspath{ {./Figures/} }
\usepackage{caption}
\usepackage{color, colortbl}
\definecolor{Gray}{gray}{0.9} 

\usepackage{booktabs} 
\usepackage{array} 
\usepackage{paralist} 
\usepackage{verbatim} 
\usepackage{subfig} 
\usepackage{bm}
\usepackage{amsmath}
\usepackage{mathtools}
\usepackage{amsthm}
\usepackage{amssymb}
\usepackage{longtable}
\usepackage{xcolor}
\usepackage{fancyvrb}
\usepackage{multirow}
\usepackage{etoolbox}
\usepackage{float}
\usepackage[colorlinks, citecolor=blue, urlcolor=blue, linkcolor=blue]{hyperref}
\usepackage[round]{natbib}
\usepackage{enumitem}
\usepackage{listings}
\usepackage{arydshln}

\let\bbordermatrix\bordermatrix
\patchcmd{\bbordermatrix}{8.75}{4.75}{}{}
\patchcmd{\bbordermatrix}{\left(}{\left[}{}{}
\patchcmd{\bbordermatrix}{\right)}{\right]}{}{}

\theoremstyle{definition}

\theoremstyle{remark}

\theoremstyle{plain}

\DeclareMathAlphabet{\mathpzc}{OT1}{pzc}{m}{it}

\usepackage{fancyhdr} 
\usepackage{sectsty}
\allsectionsfont{\sffamily\mdseries\upshape} 

\usepackage[nottoc,notlof,notlot]{tocbibind} 
\usepackage[titles,subfigure]{tocloft} 

\definecolor{mygreen}{rgb}{0,0.6,0}
\definecolor{mygray}{rgb}{0.5,0.5,0.5}
\definecolor{mediumseagreen}{rgb}{0.24, 0.7, 0.44}

\lstdefinestyle{MyRprograms}{
	language = R,
	basicstyle=\ttfamily\footnotesize,           
	otherkeywords={}, 
	numberstyle=\tiny\color{mygray},  
	stepnumber=1,                   
	numbersep=5pt,                  
	backgroundcolor=\color{white},      
	showspaces=false,               
	showstringspaces=false,         
	showtabs=false,                 
	frame=single,                   
	rulecolor=\color{black},        
	tabsize=2,                      
	captionpos=b,                   
	breaklines=true,                
	breakatwhitespace=false,        
	keywordstyle={},          
	commentstyle=\color{mediumseagreen},       
	stringstyle=\color{mygray},         
	escapeinside={},            
	morekeywords={byrow},              
	lineskip = {-1.5pt} 
}

\newcommand{\bx}{\boldsymbol{x}}

\makeatletter
\newcommand{\vast}{\bBigg@{4}}
\newcommand{\Vast}{\bBigg@{5}}
\makeatother

\graphicspath{ {Figures/} }

\begin{document}

\title{Confidence intervals in general regression models that utilize uncertain prior information}

\author{Paul Kabaila and Nishika Ranathunga}

\maketitle

\begin{center}
    \textbf{Abstract}
\end{center}
We consider a general regression model, without a scale parameter. Our aim is to construct a confidence interval for a scalar parameter of interest $\theta$ that utilizes the uncertain prior information that a distinct scalar parameter $\tau$ takes the specified value $t$. This confidence interval should have good coverage properties.
It should also have scaled expected length, where the scaling is with respect to the usual confidence interval, that (a) is substantially less than 1 when the prior information is correct, (b) has a maximum value that is not too large and (c) is close to 1 when the data and prior information are highly discordant. 
The asymptotic joint distribution of the maximum likelihood estimators $\theta$ and $\tau$ is similar to the joint distributions of these estimators 
in the particular case of a linear regression with normally distributed errors having known variance. This similarity is used to construct a confidence interval with the desired properties by using the confidence interval, computed using the \texttt{R} package 
\texttt{ciuupi}, that utilizes the uncertain prior information in this particular linear regression case.
An important practical application of this confidence interval is to 
a quantal bioassay carried out to compare two similar compounds. In this context,
the uncertain prior information is that the hypothesis of ``parallelism'' holds.
We provide extensive numerical results that illustrate the properties of this confidence interval in this context. 

\newpage

\section{Introduction}

Uncertain prior information about the values of the parameters of a model may result from previous experience with similar data sets and/or expert opinion and scientific background. We say that a frequentist confidence region for the parameter of interest utilizes the uncertain prior information if it has good coverage properties and has scaled expected volume, where the scaling is with respect to the usual confidence region, that (a) is substantially less than 1 when the prior information is correct, (b) has a maximum value that is not too large and (c) is close to 1 when the data and prior information are highly discordant.

Such regions include (a)
confidence regions for the multivariate normal mean that dominate the usual confidence region (\citeauthor{CasellaHwang2012}, \citeyear{CasellaHwang2012}),
(b) 
confidence intervals for the normal variance that dominate the usual confidence interval (\citeauthor{MaataCasella1990}, \citeyear{MaataCasella1990}), (c) the confidence regions
constructed by 
\cite{YuHoff2018} and \cite{HoffYu2019} using an extension of the ``tail method'' described by \cite{PuzaONeill2006} 
and (d) the confidence regions constructed 
by \cite{FarchioneKabaila2008}, \cite{KabailaGiri2009JSPI}, \cite{KabailaGiri2014}, \cite{KabailaTissera2014}, \cite{AbeysekeraKabaila2017} and \cite{MainzerKabaila2019}
using numerical nonlinear constrained optimization.  

We consider a general regression model, without a scale parameter. 
An example of such a model is a generalized linear model with binomial responses and canonical link function. Our aim is to construct a confidence interval for a scalar parameter of interest $\theta$ that utilizes the uncertain prior information that a distinct scalar parameter $\tau$ takes the specified value $t$. 
In Section 3, we introduce a \textsl{local asymptotic framework}, similar to the ``local misspecification framework'' of \cite{HjortClaeskens2003}, which is then used to define the \textsl{local coverage probability} and the \textsl{local scaled expected length} of a confidence interval for $\theta$.
We seek to construct a confidence interval for $\theta$ with the following characteristics. Firstly, it has endpoints that are smooth functions of the data. Secondly, it has \textsl{local coverage probability} that is close
to $1 - \alpha$. Thirdly, it has \textsl{local scaled expected length} that (a)
is substantially less than 1 when the prior information that $\tau = t$ is correct, (b) has a maximum value that is not too large and (c) approaches 1 for large $|\tau - t|$.

An important practical application of our work is to a quantal bioassay carried out to compare two similar compounds.  Let 
$\theta$ be a scalar measure of the difference between these compounds. A detailed description of this type of bioassay is given in Section 4. In this context, we commonly have uncertain prior information that the hypothesis of ``parallelism'' holds. This hypothesis can be expressed in the form $\tau$ takes the specified value $t$. Therefore our aim is to find a confidence interval for $\theta$ that utilizes this uncertain prior information. In Section 5, we provide extensive numerical results that illustrate the properties of this confidence interval.

Suppose that the distribution of the response vector, for given values of the explanatory variables, is determined by the unknown parameter vector $\bm{\beta}$.
Also 
suppose that the scalar parameter of interest $\theta = g(\bm{\beta})$ and  that the scalar parameter $\tau = h(\bm{\beta})$. 
Let $\widehat{\bm{\beta}}$
denote the maximum likelihood estimator of $\bm{\beta}$. Also let
$\widehat{\theta} = g(\widehat{\bm{\beta}})$ and 
$\widehat{\tau} = h(\widehat{\bm{\beta}})$.
The asymptotic bivariate normal distribution of 
$\big(\widehat{\theta}, \widehat{\tau} \big)$ is similar to the bivariate normal distribution of $\big(\widehat{\theta}, \widehat{\tau} \big)$
in the following particular case.
\begin{enumerate}
    \item[]
    \underline{Particular Case L} \newline
The regression model is linear and has independent and identically normally distribution random errors with known error variance. The functions $g$ and $h$ are linear. 
\end{enumerate}
 In this particular case, the \texttt{R} package \texttt{ciuupi} can be used to construct a confidence interval for $\theta$ that utilizes the uncertain prior information that $\tau = t$. We use this confidence interval to construct the  confidence interval with the desired properties in the general regression context, based on this similarity of the bivariate distributions. This 
 similarity can be expressed either in terms of Wald statistics or signed root likelihood ratio (SRLR) statistics. We have found that expressing this similarity in terms of SRLR statistics leads to the confidence interval in the general regression context having better performance than
when we express this similarity in terms of Wald statistics.

\section{The confidence interval that utilizes uncertain prior information in linear regression with known error variance
} 
\label{LinearRegression}

Consider the linear regression model
\begin{equation}
\label{OrigLinRegress}
\bm{y} = \bm{X} \bm{\beta} + \bm{\varepsilon},
\end{equation}
where $\bm{y}$ is a random $n$-vector of responses, $\bm{X}$ is a known $n \times p$ matrix with linearly independent columns, $\bm{\beta}$ is an unknown parameter $p$-vector and $\bm{\varepsilon} \sim N(\bm{0}, \, \sigma^2 \, \bm{I})$, where $\sigma^2$ is known.  Suppose that the parameter of interest is $\theta = \bm{a}^{\top} \bm{\beta}$, where $\bm{a}$ is a specified nonzero $p$-vector.  
Let $\tau = \bm{c}^{\top} \bm{\beta}$, where $\bm{c}$ is a specified nonzero $p$-vector that is linearly independent of $\bm{a}$. Suppose that we have uncertain prior information
that $\tau = t$, where $t$ is a specified number (commonly $t=0$). 
Our aim is to construct a CI for $\theta$, with minimum coverage probability $1 - \alpha$, that utilizes this uncertain prior information.

Let $\widehat{\bm{\beta}} = (\bm{X}^{\top} \bm{X})^{-1} \, \bm{X}^{\top} \, \bm{y}$, the least squares estimator of $\bm{\beta}$. Then $\widehat{\theta} = \bm{a}^{\top} \widehat{\bm{\beta}}$ and $\widehat{\tau} = \bm{c}^{\top} \widehat{\bm{\beta}}$ are the least squares estimators of $\theta$ and $\tau$, respectively.  
Note that
$\text{var}(\widehat{\theta}) 
= \sigma^2 \, \bm{a}^{\top}(\bm{X}^{\top}\bm{X})^{-1}\bm{a}$,
$\text{var}(\widehat{\tau}) = \sigma^2 \, \bm{c}^{\top}(\bm{X}^{\top} \bm{X})^{-1} \bm{c}$
and $\text{cov}(\widehat{\theta}, \widehat{\tau}) 
= \sigma^2 \, \bm{a}^{\top}(\bm{X}^{\top}\bm{X})^{-1}\bm{c}$,  
which are known quantities. Hence 
\begin{equation}
\label{RhoLinRegression}
\rho 
= \text{corr} \big(\widehat{\theta}, \widehat{\tau} \big)	
= \frac{\text{cov}(\widehat{\theta}, \widehat{\tau})}{\big(\text{var}(\widehat{\theta})  \, \text{var}(\widehat{\tau})\big)^{1/2}}
\end{equation}
is also known.

Our first step in the description of the CI for $\theta$ that utilizes the uncertain prior information is to reduce the data to 
$\big(\widehat{\theta}, \widehat{\tau}\big)$. A justification for this 
data reduction is provided by the change of parametrization described in Section 4 of the Supplementary Material for \cite{KabailaWelshAbeysekera2016} with $t = 0$.
Observe that 
\begin{align}
\label{JointDistThetahatTauhat}
\left[ {\begin{array}{c}
	\widehat{\theta} \\
	\widehat{\tau}
	\end{array} } \right] \sim N \left(
\left[ {\begin{array}{c}
	\theta \\
	\tau
	\end{array} } \right],
\left[ {\begin{array}{cc}
	\text{var}(\widehat{\theta})   & \text{cov}(\widehat{\theta}, \widehat{\tau}) 
	\\
	\text{cov}(\widehat{\theta}, \widehat{\tau})  & \text{var}(\widehat{\tau})
	\end{array} } \right]
\right).
\end{align}
Let $[a \pm w]$ denote the interval $[a - w, a + w]$ ($w > 0$).
The usual $1-\alpha$ confidence interval for $\theta$ is
\begin{equation}
\label{UsualCI_ciuupi}
\text{I} = \left[\widehat{\theta} \pm z_{1 - \alpha/2} \, \big(\text{var}(\widehat{\theta}) \big)^{1/2} \right],
\end{equation}
where the qunatile $z_p$ is defined by $P(Z \le z_p) = p$ for $Z \sim N(0,1)$.

The confidence interval for $\theta$ computed by the \texttt{R} package \texttt{ciuupi}, with minimum coverage probability $1 - \alpha$, that utilizes the uncertain prior information that $\tau = t$, has the form 
\begin{equation}
\label{eqn_ciuupi}
\text{CI}(b, s) 
= \left[ \widehat{\theta} 
- \big(\text{var}(\widehat{\theta}) \big)^{1/2} \, 
b \left( \frac{\widehat{\tau} - t}{ \big(\text{var}(\widehat{\tau})\big)^{1/2}}\right)  
\pm \big(\text{var}(\widehat{\theta}) \big)^{1/2} \, s \left( \frac{\widehat{\tau} - t}{ \big(\text{var}(\widehat{\tau})\big)^{1/2}}\right) \right],
\end{equation}
where $b: \mathbb{R} \to \mathbb{R}$ is an odd continuous function and $s: \mathbb{R} \to \mathbb{R}$ is an even continuous function.  
In addition, $b(x) = 0$ and $s(x) = z_{1 - \alpha/2}$ for all $|x| \geq 6$.
Define the scaled expected length of $\text{CI}(b,s)$ to be $E(\text{length of} \ \text{CI}(b,s)) / (\text{length of I})$.
The \texttt{R} package \texttt{ciuupi} computes the functions $b$ and $s$ such that $\text{CI}(b, s)$ has the following properties. It has minimum coverage probability $1 - \alpha$ and scaled expected length for $\tau = t$ that is as small as possible, subject to an upper bound on its maximum value. \cite{MainzerKabaila2019} describe this computation in full detail. The functions $b$ and $s$ that are computed by 
\texttt{ciuupi} are determined by $\rho$ and $1 - \alpha$. We denote them by $b_{\rho}$ and $s_{\rho}$, respectively, so that the confidence interval 
computed by the \texttt{R} package \texttt{ciuupi} is $\text{CI}(b_{\rho}, s_{\rho})$.

Let $\gamma = (\tau - t) / \big(\text{var}(\widehat{\tau})\big)^{1/2}$.
The coverage probability of $\text{CI}(b_{\rho},s_{\rho})$ is a function of $\gamma$, for given $\rho$. We denote this function by $CP(\gamma; \rho)$. 
Note that $CP(\gamma; \rho)$  is an even function of $\gamma$ for every given $\rho$ and an even function of $\rho$ for every given $\gamma$. 
For later reference, we make the very simple observation that the scaled expected length of $\text{CI}(b_{\rho},s_{\rho})$ is 
\begin{equation}
\label{DefSELciuupi}
\frac{E \big(\text{length of} \ \text{CI}(b_{\rho}, s_{\rho}) \big)}{ \big(\text{length of I} \big)} 
= E \left(\frac{\text{length of} \ \text{CI}(b_{\rho}, s_{\rho})}{\text{length of I computed from the same data } }\right).
\end{equation}
The scaled expected length of $\text{CI}(b_{\rho},s_{\rho})$ is a function of $\gamma$, for given $\rho$. We denote this function by $SEL(\gamma; \rho)$. 
Note that  $SEL(\gamma; \rho)$ is an even function of $\gamma$ for every given $\rho$ and an even function of $\rho$ for every given $\gamma$.

For later reference, we note that the confidence interval $\text{CI}(b_{\rho},s_{\rho})$ can be expressed 
in terms of likelihood functions as follows.
Let $l(\theta, \tau)$ denote the log-likelihood function based on 
$(\widehat{\theta}, \widehat{\tau})$. The distribution of 
$(\widehat{\theta}, \widehat{\tau})$ is given by \eqref{JointDistThetahatTauhat}.
Let $\widehat{\tau}_{\theta}$ denote the value of $\tau$ that maximizes 
$l(\theta, \tau)$ with respect to $\tau$, for given $\theta$. 
Now define the SRLR statistic 
\begin{equation*}
\text{r}_1(\theta^{\prime})
= \text{sign} (\widehat{\theta} - \theta^{\prime}) \sqrt{2 \, \left(l\big(\widehat{\theta}, \widehat{\tau}\big) - l\big(\theta^{\prime}, \widehat{\tau}_{\theta^{\prime}}\big)\right)}.
\end{equation*}
Let $\widehat{\theta}_{t}$ denote the value of $\theta$ that maximizes
$l(\theta, \tau)$ with respect to $\theta$, for $\tau = t$. Now define the SRLR statistic
\begin{equation*}
\text{r}_2
= \text{sign} (\widehat{\tau} - t) \sqrt{2 \, \left(l\big(\widehat{\theta}, \widehat{\tau}\big) - l\big(\widehat{\theta}_t, t\big)\right)}.
\end{equation*}
For notational convenience, let $v_{\theta} = \text{var}(\widehat{\theta})$ and $v_{\tau} = \text{var}(\widehat{\tau})$.
Clearly, $\text{CI}(b_{\rho}, s_{\rho})$ is equal to 
\begin{equation*}
\left \{ \theta^{\prime} \in \mathbb{R}: b_{\rho} \left( \frac{\widehat{\tau} - t}{v_{\tau}^{1/2}}\right)  
- s_{\rho} \left( \frac{\widehat{\tau} - t}{v_{\tau}^{1/2}}\right) \leq \frac{\widehat{\theta}- \theta^{\prime}}{v_{\theta}^{1/2}} \leq
b_{\rho} \left( \frac{\widehat{\tau} - t}{v_{\tau}^{1/2}}\right)  
+ s_{\rho} \left( \frac{\widehat{\tau} - t}{v_{\tau}^{1/2}}\right) \right \}.
\end{equation*}
It may be shown that 
\begin{equation*}
\frac{\widehat{\theta} - \theta^{\prime}}{v_{\theta}^{1/2}}
= \text{r}_1(\theta^{\prime}) 
\quad \text{and} \quad 
\frac{\widehat{\tau} - t}{v_{\tau}^{1/2}} = \text{r}_2.
\end{equation*}
Hence
the confidence interval $\text{CI}(b_{\rho}, s_{\rho})$ is given by 
\begin{equation}
\label{ciuupi_LikelihoodExpression}
\Big\{\theta^{\prime} \in \mathbb{R}:
b_{\rho} \left( \text{r}_2\right)  
- s_{\rho} \left( \text{r}_2\right) \leq \text{r}_1(\theta^{\prime}) \leq
b_{\rho} \left( \text{r}_2\right)  
+ s_{\rho} \left( \text{r}_2\right) \Big\}.
\end{equation}

\section{Asymptotic results for a general regression model, without a scale parameter}
\label{GenRegressModel}


In this section, we consider a general regression model, without a scale parameter. 
Using the well-known asymptotic distribution of the maximum likelihood estimator, we derive an asymptotic distribution that is analogous to the distribution \eqref{JointDistThetahatTauhat}, which is for the linear regression model considered in the previous section. 

We consider a general regression model with response vector 
$\bm{y} = (y_1, \dots, y_n)$. 
The random variables $y_1, \dots, y_n$ are independent
and $y_i$ has pmf or pdf (as as appropriate) $f_i(y \, | \, \bx_i, \bm{\beta})$, where $\bm{\beta} = (\beta_1, \dots, \beta_p)$ is an unknown parameter $p$-vector, which belongs to the open set ${\cal B}$, and
$\bm{x}_i$ a vector of explanatory variables of given dimension ($i=1, \dots, n$).
Suppose that the scalar parameter of interest $\theta = g(\bm{\beta})$,
where $g: \mathbb{R}^p \rightarrow \mathbb{R}$ is a sufficiently smooth function.
Also suppose that the parameter $\tau = h(\bm{\beta})$,
where $h: \mathbb{R}^p \rightarrow \mathbb{R}$ is a sufficiently smooth function. 
Let $\partial g(\bm{\beta}) / \partial \bm{\beta}$ denote the row $p$-vector 
with $i$th component 
$\partial g(\bm{\beta}) / \partial \beta_i$ ($i = 1, \dots, p$).
Suppose that $\partial g(\bm{\beta}) / \partial \bm{\beta}$
and $\partial h(\bm{\beta}) / \partial \bm{\beta}$ are linearly independent $p$-vectors, for all $\bm{\beta} \in {\cal B}$.
Finally, suppose that we have uncertain prior information that $\tau = t$, where $t$ is a specified number.

Let $I(\bm{\beta})$ denote the Fisher information matrix. In other words, $I(\bm{\beta})$
is the $p \times p$ matrix with $(i,j)$th element
\begin{equation*}
-\sum_{i=1}^n E\left( \dfrac{\partial^2  \log f_i(y_i \, | \, \bx_i ; \bm{\beta})}{\partial \beta_i \; \partial \beta_j} \right).
\label{FisherInfoMatrix}
\end{equation*}
We suppose that $I(\bm{\beta})$ is nonsingular for all 
$\bm{\beta} \in {\cal B}$. For convenience, we do not make the dependence of this matrix on $n$ explicit in the notation.
We also suppose that 
$n^{-1} I(\bm{\beta})$ converges to a finite nonsingular matrix as 
$n \rightarrow \infty$, for each $\bm{\beta} \in {\cal B}$.

Denote the maximum likelihood estimator of $\bm{\beta}$ by $\widehat{\bm{\beta}}$. 
Under the appropriate regularity conditions, 
\begin{equation*}
n^{1/2} \Big(\widehat{\bm{\beta}} - \bm{\beta} \Big) \; \stackrel{approx}{\sim} \,
N \left(\bm{0}, \Big(n^{-1} I(\bm{\beta}) \Big)^{-1}\right),
\end{equation*}
for large $n$. We use the following shorthand for this large sample distribution
\begin{equation}
\label{AsymptDistBetaHat}
\widehat{\bm{\beta}}  \; \stackrel{asympt}{\sim} \,
N \left(\bm{\beta}, \Big(I(\bm{\beta}) \Big)^{-1}\right).
\end{equation}
Let $\widehat{\theta} = g(\widehat{\bm{\beta}})$ and 
$\widehat{\tau} = h(\widehat{\bm{\beta}})$ denote the maximum likelihood estimators of $\theta$ and $\tau$, respectively. 
Similarly to Section \ref{LinearRegression}, our first step in the description of the CI for $\theta$ that utilizes the uncertain prior information is to reduce the data to 
$\big(\widehat{\theta}, \widehat{\tau}\big)$. 

By the mean value theorem,
\begin{align*}
\widehat{\theta} - \theta 
\approx \frac{\partial g(\bm{\beta})}{\partial \bm{\beta}} \, \big(\widehat{\bm{\beta}} - \bm{\beta} \big) \ \ 
\text{and}\ \ \widehat{\tau} - \tau 
\approx \frac{\partial h(\bm{\beta})}{\partial \bm{\beta}} \, \big(\widehat{\bm{\beta}} - \bm{\beta} \big).
\end{align*}
Therefore
\begin{align}
\label{AsympDist}
\begin{bmatrix}
\widehat{\theta}
\\
\widehat{\tau} 
\end{bmatrix}
\stackrel{asympt}{\sim} \,
N \left(\begin{bmatrix} 
\theta
\\
\tau
\end{bmatrix} , 
\begin{bmatrix} 
\text{avar}\big(\widehat{\theta}; \bm{\beta} \big)
& \text{acov}\big(\widehat{\theta}, \widehat{\tau}; \bm{\beta}\big)
\\
\text{acov}\big(\widehat{\theta}, \widehat{\tau}; \bm{\beta}\big)
& \text{avar}\big(\widehat{\tau}; \bm{\beta} \big)
\end{bmatrix} 
\right),
\end{align}
where $\text{avar}\big(\widehat{\theta}; \bm{\beta} \big)$ denotes the asymptotic variance of $\widehat{\theta}$, $\text{acov}\big(\widehat{\theta}, \widehat{\tau}; \bm{\beta}\big)$
denotes the asymptotic covariance of $\widehat{\theta}$ and $\widehat{\tau}$,
\begin{align*}
\text{avar}\big(\widehat{\theta}; \bm{\beta} \big)
&= \frac{\partial g(\bm{\beta})}{\partial \bm{\beta}} \,
\big(I(\bm{\beta}) \big)^{-1}  
\left(\frac{\partial g(\bm{\beta})}{\partial \bm{\beta}} \right)^{\top},
\\
\text{avar}\big(\widehat{\tau}; \bm{\beta} \big)
&= \frac{\partial h(\bm{\beta})}{\partial \bm{\beta}} \,
\big(I(\bm{\beta}) \big)^{-1}  
\left(\frac{\partial h(\bm{\beta})}{\partial \bm{\beta}} \right)^{\top},
\\
\text{and} \ \ \ \ \text{acov}\big(\widehat{\theta}, \widehat{\tau}; \bm{\beta}\big)
&= \frac{\partial g(\bm{\beta})}{\partial \bm{\beta}} \,
\big(I(\bm{\beta}) \big)^{-1}  
\left(\frac{\partial h(\bm{\beta})}{\partial \bm{\beta}} \right)^{\top}.
\end{align*}
Similarly to \eqref{RhoLinRegression}, let
\begin{equation}
\label{RhoGenRegression}
\rho (\bm{\beta})
= \frac{\text{acov}\big(\widehat{\theta}, \widehat{\tau}; \bm{\beta}\big)}{\Big(\text{avar}\big(\widehat{\theta}; \bm{\beta} \big) \; \text{avar}\big(\widehat{\tau}; \bm{\beta} \big)\Big)^{1/2}}.
\end{equation}

\subsection{Analogues of $\text{I}$ and 
	$\text{CI}(b_{\rho}, s_{\rho})$ based on Wald statistics}

In this section, we describe  analogues of the confidence intervals $\text{I}$ and 
$\text{CI}(b_{\rho}, s_{\rho})$ based on Wald statistics.
The analogue of the confidence interval $\text{I}$, given by \eqref{UsualCI_ciuupi} and based on the assumption that $\big(\widehat{\theta} - \theta\big) \big / \big(\text{avar}\big(\widehat{\theta}; \widehat{\bm{\beta}}\big) \big)^{1/2}$
has approximately an $N(0,1)$ distribution, is $\text{I}_{\text W}(\bm{y}; \alpha)$, where
\begin{equation*}
\text{I}_{\text W}(\bm{y};c) 
= \left[\widehat{\theta} \pm 
z_{1 - c/2} \left(\text{avar}\big(\widehat{\theta}; \widehat{\bm{\beta}}  \big) \right)^{1/2} \right],
\end{equation*}
with $0 < c < 1/2$.

Let $\text{ACI}_{\text W}\big(\bm{\beta}\big)$ denote the interval
\begin{equation*}
\left[ \widehat{\theta} 
- \big(\text{avar}\big(\widehat{\theta}; \bm{\beta} \big) \big)^{1/2} \, 
b_{\rho(\bm{\beta})} \left( \frac{\widehat{\tau} - t}{ \big(\text{avar}\big(\widehat{\tau}; \bm{\beta} \big)\big)^{1/2}}\right)  
\pm \big(\text{avar}\big(\widehat{\theta}; \bm{\beta} \big)\big)^{1/2} \, 
s_{\rho(\bm{\beta})} \left( \frac{\widehat{\tau} - t}{ \big(\text{avar}\big(\widehat{\tau}; \bm{\beta} \big)\big)^{1/2}}\right)  \right],
\end{equation*}
where the functions $b_{\rho(\bm{\beta})}$ and $s_{\rho(\bm{\beta})}$ are the functions $b$ and $s$, respectively, computed using the \texttt{R} package \texttt{ciuupi}, with the desired minimum coverage probability $1- \alpha$ and 
$\rho = \rho(\bm{\beta})$.
We now apply the plug-in principle to obtain the confidence interval 
$\text{ACI}_{\text{W}}(\widehat{\bm{\beta}})$ for $\theta$.
This confidence interval is given by 
\begin{equation*}
\left[ \widehat{\theta} 
- \big(\text{avar}\big(\widehat{\theta}; \widehat{\bm{\beta}} \big) \big)^{1/2} \, 
b_{\rho(\widehat{\bm{\beta}})} \left( \frac{\widehat{\tau} - t}{ \big(\text{avar}\big(\widehat{\tau}; \widehat{\bm{\beta}} \big)\big)^{1/2}}\right)  
\pm \big(\text{avar}\big(\widehat{\theta}; \widehat{\bm{\beta}} \big)\big)^{1/2} \, 
s_{\rho(\widehat{\bm{\beta}})} \left( \frac{\widehat{\tau} - t}{ \big(\text{avar}\big(\widehat{\tau}; \widehat{\bm{\beta}} \big)\big)^{1/2}}\right)  \right],
\end{equation*}
Note that $\big(\widehat{\tau} - t\big) \big / \big(\text{avar}\big(\widehat{\tau}; \widehat{\bm{\beta}}\big) \big)^{1/2}$ is the 
Wald test statistic for testing the null hypothesis $H_0: \tau =t$ against the alternative hypothesis $H_A: \tau \ne t$.
The similarity between 
the bivariate normal distribution 
\eqref{JointDistThetahatTauhat}
and asymptotic bivariate normal distribution
\eqref{AsympDist} suggests that 
$\text{ACI}_{\text W}\big(\widehat{\bm{\beta}}\big)$ will have coverage probability approximately equal to $1 - \alpha$ and the desired expected length properties.
To summarize, the analogues of the confidence intervals $\text{I}$ and $\text{CI}(b_{\rho}, s_{\rho})$, based on Wald statistics, are 
$\text{I}_{\text W}(\bm{y};\alpha)$
and $\text{ACI}_{\text W}\big(\widehat{\bm{\beta}}\big)$, respectively.

\subsection{Analogues of $\text{I}$ and 
	$\text{CI}(b_{\rho}, s_{\rho})$  based on likelihood functions}

There is some evidence that likelihood based methods lead to better hypothesis tests and confidence intervals than Wald based methods, see e.g. 
\cite{MeekerEscobar1995}, \citeauthor{Cox2006} (\citeyear{Cox2006}, p.117--118),
\cite{Pawitan2000},  \citeauthor{YoungSmith2005} (\citeyear{YoungSmith2005}, p.137)
and \cite{Pawitan2013}. For this reason, in this subsection, we describe 
analogues of $\text{I}$ and 
$\text{CI}(b_{\rho}, s_{\rho})$  based on likelihood functions.

\subsubsection{Profile likelihood confidence interval for $\theta$}

The profile likelihood confidence interval for $\theta$, with nominal coverage $1 - \alpha$, is the likelihood-based analogue of the confidence interval $\text{I}$.
Let $\ell(\bm{\beta} \, | \, \bm{y})$ denote the log-likelihood function for the general regression model, without a scale parameter, described in Section \ref{GenRegressModel}. To compute the profile likelihood confidence interval for $\theta$, with nominal coverage $1 - \alpha$, we invert a family of hypothesis tests. 
We test the null hypothesis $H_0 : \theta = \theta^{\prime}$ against the alternative hypothesis $H_A : \theta \neq \theta^{\prime}$ using the SRLR statistic
\begin{equation}
r_1(\theta^{\prime}\, | \, \bm{y}) = 
\text{sign} \big(\widehat{\theta} - \theta^{\prime} \big) \sqrt{2 \, \left(\ell\big(\widehat{\bm{\beta}} \, \big| \, \bm{y}\big) - \ell\big(\widehat{\bm{\beta}}(\theta^{\prime}; \theta) \, \big| \, \bm{y} \big)\right)},
\label{SRLRstatisticTheta}
\end{equation}
where  
$\widehat{\bm{\beta}}(\theta^{\prime}; \theta)$ maximises 
$\ell(\bm{\beta} \, | \, \bm{y})$ with respect to $\bm{\beta}$, subject to the constraint that $g(\bm{\beta}) = \theta^{\prime}$.
Suppose that we accept $H_0$ if and only if 
$-z_{1 - c/2} \leq r_1(\theta^{\prime} \, | \, \bm{y}) \leq z_{1 - c/2}$,
where $0 < c < 1/2$.
The confidence set, with nominal coverage $1 - c$ and found by inverting the family of hypothesis tests obtained as are vary over $\theta^{\prime} \in \mathbb{R}$, is
\begin{equation}
{\cal S}_{\text{PL}}(\bm{y}) = \left\{\theta^{\prime} \in \mathbb{R} : -z_{1 - c/2} \leq r_1(\theta^{\prime} \, | \, \bm{y}) \leq z_{1 - c/2}\right\}.
\label{ConfidenceSetTheta}
\end{equation}
Define the profile likelihood confidence interval, with nominal coverage $1 - \alpha$, as follows. This confidence interval, denoted by $\text{I}_{\text{L}}(\bm{y}; c)$,
has lower endpoint $\inf\big({\cal S}_{\text{PL}}(\bm{y}) \big)$ and upper endpoint
$\sup\big({\cal S}_{\text{PL}}(\bm{y}) \big)$. When $r_1(\theta^{\prime}\, | \, \bm{y})$ is a decreasing function of $\theta^{\prime}$, 
 $\text{I}_{\text{L}}(\bm{y};c) = \left[\widehat{\theta}_l, \widehat{\theta}_u\right]$, where $\widehat{\theta}_l$ and $\widehat{\theta}_u$ are the solutions for $\theta^{\prime}$ of 
\begin{equation}
  r_1(\theta^{\prime} \, | \, \bm{y}) = z_{1 - c/2} 
\quad \text{and} \quad
 r_1(\theta^{\prime} \, | \, \bm{y}) = - z_{1 - c/2} ,
\label{ProfLikCI}
\end{equation}
respectively.
To summarize, the analogue of the confidence interval $\text{I}$, based on likelihood functions, is $\text{I}_{\text{L}}(\bm{y};\alpha)$.

\subsubsection{Likelihood-based analogue of $\text{CI}(b_{\rho}, s_{\rho})$}

The SRLR test statistic for testing $H_0 : \tau = t$ against the alternative hypothesis $H_A : \tau \neq t$ is
\begin{equation*}
r_2(\bm{y})
= \text{sign} (\widehat{\tau} - t) 
\sqrt{2 \, \left(\ell\big(\widehat{\bm{\beta}} \, | \, \bm{y}\big) - \ell\big(\widehat{\bm{\beta}}(t; \tau) \, | \, \bm{y}\big)\right)},
\end{equation*}
where  
$\widehat{\bm{\beta}}(t; \tau)$ maximises 
$\ell(\bm{\beta} \, | \, \bm{y})$ with respect to $\bm{\beta}$, subject to the constraint that $h(\bm{\beta}) = t$.
The likelihood based confidence set for $\theta$, with nominal coverage $1 - \alpha$, that is analogous to \eqref{ciuupi_LikelihoodExpression} is
\begin{equation}
\label{ACI_L}
{\cal S}_{\text{A}}(\bm{y}) 
= \Big\{\theta^{\prime} \in \mathbb{R}:
b_{\rho(\widehat{\bm{\beta}})} \big(r_2(\bm{y}) \big) 
- s_{\rho(\widehat{\bm{\beta}})}
\big(r_2(\bm{y})\big) \leq r_1(\theta^{\prime} \, | \, \bm{y}) \leq b_{\rho(\widehat{\bm{\beta}})}
\big(r_2(\bm{y})\big) + s_{\rho(\widehat{\bm{\beta}})} \big(r_2(\bm{y})\big) \Big\}.
\end{equation}
Define the confidence interval $\text{ACI}_{\text L}(\bm{y})$, with nominal coverage $1 - \alpha$, as follows. This confidence interval has lower endpoint
$\inf \big({\cal S}_{\text{A}}(\bm{y})\big)$ and upper endpoint $\sup \big({\cal S}_{\text{A}}(\bm{y})\big)$. When $r_1(\theta^{\prime}\, | \, \bm{y})$ is a decreasing function of $\theta^{\prime}$, 
 $\text{ACI}_{\text L}(\bm{y}) = \left[\widetilde{\theta}_l, \widetilde{\theta}_u\right]$, where $\widetilde{\theta}_l$ and $\widetilde{\theta}_u$ are the solutions for $\theta^{\prime}$ of 
\begin{equation}
\label{ACI_L_CI}
 r_1(\theta^{\prime} \, | \, \bm{y}) 
 = b_{\rho(\widehat{\bm{\beta}})} \big(r_2(\bm{y})\big) + s_{\rho(\widehat{\bm{\beta}})} \big(r_2(\bm{y})\big) 
 \quad \text{and} \quad 
  r_1(\theta^{\prime} \, | \, \bm{y}) 
  = b_{\rho(\widehat{\bm{\beta}})} \big(r_2(\bm{y})\big) - s_{\rho(\widehat{\bm{\beta}})} \big(r_2(\bm{y})\big),
\end{equation}
respectively. 
To summarize, the analogue of the confidence interval $\text{CI}(b_{\rho}, s_{\rho})$, based on likelihood functions, is $\text{ACI}_{\text L}(\bm{y})$.

\subsection{Assessment of the coverage probability of a confidence interval}

How should we assess
the coverage probability of the confidence interval
$\text{ACI}_{\text W}\big(\widehat{\bm{\beta}}\big)$, which has nominal coverage 
$1 - \alpha$? For the sake of concreteness, suppose that the response $y_i \sim \text{Binomial}(N_i, \psi_i)$, with $N_i$ given ($i = 1, \dots n$).
 Let 
$\text{logit}(x) = \log(x / (1 - x))$, for $0 < x < 1$. Also suppose that 
$\text{logit}(\psi_i) = \sum _{j = 1}^p x_{ij} \beta_j$, where the $x_{ij}$ are explanatory variables taking positive values and $\beta_1, \dots, \beta_p$ are unknown parameters. The coverage probabilty 
$P_{\bm{\beta}} \big(\theta \in \text{ACI}_{\text W}\big(\widehat{\bm{\beta}}\big)\big)$ will take values far below $1 - \alpha$ for extreme values of $\bm{\beta}$, such as when $\beta_1, \dots, \beta_p$ all have the same sign and $|\beta_1|, \dots, |\beta_p|$ are all large.
In fact, the infimum over $\bm{\beta} \in \mathbb{R}^p$ of $P_{\bm{\beta}} \big(\theta \in \text{ACI}_{\text W}\big(\widehat{\bm{\beta}}\big)\big)$ is 0. 
We expect that such extreme values of $\bm{\beta}$ are unlikely to occur
in practice, 
so that this assessment of the coverage probability of $\text{ACI}_{\text W}\big(\widehat{\bm{\beta}}\big)$ is unduly conservative.

\subsubsection{Definition of the \textsl{local minimum coverage probability}, for given 
	$\widetilde{\boldsymbol{\beta}}$}

Let $\widetilde{\bm{\beta}}$ be a given value that satisfies $h(\widetilde{\bm{\beta}}) = t$.
We deal with the choice of $\widetilde{\bm{\beta}}$ in the next subsubsection.
Consider the straight line consisting of the values of $\bm{\beta}$ satisfying
\begin{equation}
\label{beta_straight_line}
\bm{\beta}
= \widetilde{\bm{\beta}} 
+ \kappa \, \big(\partial h(\widetilde{\bm{\beta}}) \big/ \partial \bm{\beta} \big)^{\top},
\end{equation}
where 
$\partial h(\widetilde{\bm{\beta}}) \big/ \partial \bm{\beta}$ denotes
the row vector with $i$th element 
$\partial h(\widetilde{\bm{\beta}}) \big/ \partial \bm{\beta}_i$
and $\kappa \in \mathbb{R}$. 
Let $\| \cdot \|$ denote the Euclidean norm.
For given small $\| \bm{\beta} - \widetilde{\bm{\beta}} \|$, 
$|\tau - t| = \big|h(\bm{\beta}) - h(\widetilde{\bm{\beta}}) \big|$ is maximized by choosing $\bm{\beta}$ to satisfy \eqref{beta_straight_line}.

We will assess the coverage probability of $\text{ACI}_{\text W}\big(\widehat{\bm{\beta}}\big)$ for values of $\bm{\beta}$ satisfying
\eqref{beta_straight_line}
and for
\begin{equation}
\label{SegmentInfo}
\kappa =  
\frac{\big(\text{avar}(\widehat{\tau}; \widetilde{\bm{\beta}}) \big)^{1/2}}
{\big \| \partial h(\widetilde{\bm{\beta}}) \big/ \partial \bm{\beta} \big\|^2} \;
\gamma_a, \quad \text{where} \quad \gamma_a \in [-u, u]
\end{equation}
and
the chosen value of 
$u$ satisfies $1 \le u \le 10$. For the numerical illustration presented in Section 5, we have chosen $u = 2.5$. 
For the straight line segment of values of $\bm{\beta}$ satisfying
\eqref{beta_straight_line}
and \eqref{SegmentInfo},
\begin{equation*}
\big\| \bm{\beta} - \widetilde{\bm{\beta}} \big\|
= \frac{\big(\text{avar}(\widehat{\tau}; \widetilde{\bm{\beta}}) \big)^{1/2}}
{\big \| \partial h(\widetilde{\bm{\beta}}) \big/ \partial \bm{\beta} \big\|} \;
|\gamma_a|
\le 
\frac{\big(\text{avar}(\widehat{\tau}; \widetilde{\bm{\beta}}) \big)^{1/2}}
{\big \| \partial h(\widetilde{\bm{\beta}}) \big/ \partial \bm{\beta} \big\|} \;
10.
\end{equation*}
Thus the supremum, over the values of $\bm{\beta}$ satisfying
\eqref{beta_straight_line}
and \eqref{SegmentInfo}, of 
$\big\| \bm{\beta} - \widetilde{\bm{\beta}} \big\|$
converges to 0, as $n \rightarrow \infty$.
Let 
$\gamma 
= (\tau - t) \big/ \big(\text{avar}(\widehat{\tau}; \widetilde{\bm{\beta}}) \big)^{1/2}$.
Recall that $\tau = h(\bm{\beta})$ and $\widehat{\tau} = h(\widehat{\bm{\beta}})$.
Note that 
$\gamma - \gamma_a \rightarrow 0$, as $n \rightarrow \infty$.

For given $\widetilde{\bm{\beta}}$
and for $\bm{\beta}$ satisfying \eqref{beta_straight_line}
and \eqref{SegmentInfo}, the coverage probability
$P_{\bm{\beta}} \big(\theta \in \text{ACI}_{\text W}\big(\widehat{\bm{\beta}}\big)\big)$ is a function of the scalar parameter 
$\gamma_a \in [-u, u]$. For given $\widetilde{\bm{\beta}}$, we define the \textsl{local minimum coverage probability} of the confidence interval $\text{ACI}_{\text W}\big(\widehat{\bm{\beta}}\big)$
to be the minimum over the set of 
$\bm{\beta}$ satisfying \eqref{beta_straight_line}
and \eqref{SegmentInfo} of
$P_{\bm{\beta}} \big(\theta \in \text{ACI}_{\text W}\big(\widehat{\bm{\beta}}\big)\big)$.

\subsubsection{Data-based choice of $\widetilde{\boldsymbol{\beta}}$}
\label{Data_Based_Choice_Of_Beta_tilde}

Consider a given data set that is assumed to be correctly modelled by the model described at the start of Section 3. As before, let $\widehat{\bm{\beta}}$ 
denote the maximum likelihood estimate of $\bm{\beta}$. 
We choose $\widetilde{\bm{\beta}}$ to be the value of $\bm{\beta}$ that (a) satisfies
$h(\bm{\beta}) = t$ and (b) minimizes 
$\big\| \widetilde{\bm{\beta}} - \widehat{\bm{\beta}} \big\|$.
This ensures that the $\widetilde{\bm{\beta}}$ is a realistic value.

Let
\begin{equation}
\label{Beta_Star_Straight_Line}
\bm{\beta}^*
= \widetilde{\bm{\beta}} 
+ \kappa \, \big(\partial h(\widetilde{\bm{\beta}}) \big/ \partial \bm{\beta} \big)^{\top},
\end{equation}
where $\kappa$ satisfies \eqref{SegmentInfo}.
Now let $\theta^* = g(\bm{\beta}^*)$ and let $\widehat{\bm{\beta}}^*$
denote the maximum likelihood estimator of $\bm{\beta}^*$.
We assess the coverage probability 
$P_{\bm{\beta}^*} \big(\theta^* \in \text{ACI}_{\text W}\big(\widehat{\bm{\beta}}^*\big)\big)$
of the confidence interval 
$\text{ACI}_{\text W}\big(\widehat{\bm{\beta}}^*\big)$,
when the true parameter value is set to 
$\bm{\beta}^*$, by Monte Carlo simulation for each value in an equally-spaced 
grid of values of $\gamma_a \in [-u, u]$. 
These simulation results can then be used to estimate
the \textsl{local minimum coverage probability} of the confidence interval $\text{ACI}_{\text W}\big(\widehat{\bm{\beta}}^*\big)$, which is
the minimum over the set of 
$\bm{\beta}^*$ satisfying \eqref{Beta_Star_Straight_Line}, where $\kappa$ satisfies
\eqref{SegmentInfo}, of
$P_{\bm{\beta}^*} \big(\theta^* \in \text{ACI}_{\text W}\big(\widehat{\bm{\beta}}^*\big)\big)$.

Let $\widehat{\theta}^* = g(\widehat{\bm{\beta}}^*)$, $\tau^* = h(\bm{\beta}^*)$
and 
$\widehat{\tau}^* = h(\widehat{\bm{\beta}}^*)$.
It follows from the asymptotic distribution 
\eqref{AsympDist} and Slutsky's theorem that a large sample approximation to the 
distribution of $\big(\widehat{\theta}^*, \widehat{\tau}^* \big)$ is
\begin{align}
\label{LocalApproxDist}
N \left(\begin{bmatrix} 
\theta^*
\\
\tau^*
\end{bmatrix} , 
\begin{bmatrix} 
\text{avar}\big(\widehat{\theta}; \widetilde{\bm{\beta}} \big)
& \text{acov}\big(\widehat{\theta}, \widehat{\tau}; \widetilde{\bm{\beta}}\big)
\\
\text{acov}\big(\widehat{\theta}, \widehat{\tau}; \widetilde{\bm{\beta}}\big)
& \text{avar}\big(\widehat{\tau}; \widetilde{\bm{\beta}} \big)
\end{bmatrix} 
\right).
\end{align}
This distribution is obtained when we set $\text{var}(\widehat{\theta})$,
$\text{cov}(\widehat{\theta}, \widehat{\tau})$ and
$\text{var}(\widehat{\tau})$ equal to 
$\text{avar}\big(\widehat{\theta}; \widetilde{\bm{\beta}} \big)$,
$\text{acov}\big(\widehat{\theta}, \widehat{\tau}; \widetilde{\bm{\beta}}\big)$ and
$\text{avar}\big(\widehat{\tau}; \widetilde{\bm{\beta}} \big)$,
 respectively, in \eqref{JointDistThetahatTauhat}. 
 Consequently, a large sample approximation to the coverage probability 
 $P_{\bm{\beta}^*} \big(\theta^* \in \text{ACI}_{\text W}\big(\widehat{\bm{\beta}}^*\big)\big)$
 is given by 
 $CP(\gamma^*, \rho(\widetilde{\bm{\beta}}))$,
 the coverage probability of the confidence interval 
 $\text{CI}(b_{\rho(\widetilde{\bm{\beta}})}, s_{\rho(\widetilde{\bm{\beta}})})$ computed using
 \texttt{ciuupi}.

\subsection{Definition of the \textsl{local scaled expected length} of a confidence interval}
\label{Sect:DefnLocalScaledExpectedLength}

We consider the local parametric framework described in subsubsection 
\ref{Data_Based_Choice_Of_Beta_tilde}. It is within this framework that we define
the \textsl{local scaled expected length} of the confidence interval 
$\text{ACI}_{\text W}\big(\widehat{\bm{\beta}}^*\big)$, which has nominal coverage $1 - \alpha$. This definition is an analogue of the definition of the scaled expected length of the confidence interval $\text{CI}(b_{\rho}, s_{\rho})$,
as given by the right-hand side of \eqref{DefSELciuupi}. The definition of the scaled expected length, as given by the right-hand side of \eqref{DefSELciuupi},
is reasonable since the minimum coverage probabilities of 
$\text{CI}(b_{\rho}, s_{\rho})$ and $\text{I}$ are the same. However, the 
\textsl{local minimum coverage probabilities} of 
$\text{ACI}_{\text W}\big(\widehat{\bm{\beta}}^*\big)$
and $\text{I}_{\text{W}}(\bm{y}^*; \alpha)$ may not be the same. Therefore we define 
the \textsl{local scaled expected length} of 
$\text{ACI}_{\text W}\big(\widehat{\bm{\beta}}^*\big)$
to be
\begin{equation*}
E_{\bm{\beta}^*} 
\left(\frac{\text{length of} \ \text{ACI}_{\text W}\big(\widehat{\bm{\beta}}^*\big) }
{\text{length of} \ \text{I}_{\text{W}}(\bm{y}^*;\widetilde{c}) \ \text{computed from the same data}}\right),
\end{equation*}
where $\widetilde{c}$ is such that the \textsl{local minimum coverage probabilities} of $\text{ACI}_{\text W}\big(\widehat{\bm{\beta}}^*\big)$ and $\text{I}_{\text{W}}(\bm{y}^*;\widetilde{c})$
are the same.

Consider the argument given in the last paragraph of subsubsection \ref{Data_Based_Choice_Of_Beta_tilde}. This argument implies that a large sample approximation to the \textsl{local scaled expected length} of $\text{ACI}_{\text W}\big(\widehat{\bm{\beta}}^*\big)$ is given by 
$SEL(\gamma^*, \rho(\widetilde{\bm{\beta}}))$,
 the scaled expected length of the confidence interval 
 $\text{CI}(b_{\rho(\widetilde{\bm{\beta}})}, s_{\rho(\widetilde{\bm{\beta}})})$ computed using the \texttt{R} package
 \texttt{ciuupi}.


\section{Application to quantal bioassays}

We consider a quantal bioassay carried out to compare two similar compounds, labelled
A and B. This comparison is with respect to a specified dichotomous response, labelled S and not-S, for individuals that belong to a 
specified large homogeneous population. Let 
$d_1, \dots, d_m$ denote given dose levels. Now let $x_i = \log_{10}(d_i)$ for 
$i = 1, \dots, m$. Suppose that $\text{n}_1, \dots, \text{n}_m$ 
and $\text{n}_1^{\prime}, \dots, \text{n}_m^{\prime}$ are given positive integers. 

One half of the experiment consists of carrying out the following steps for each 
$i = 1, \dots, m$. Suppose that $\text{n}_i$ individuals are chosen at random from the population and given dose $d_i$ of compound A. Let $r_i$ denote the number of these individuals with response S. 
The other half of the experiment consists of carrying out the following steps for each 
$i = 1, \dots, m$. Suppose that $\text{n}_i^{\prime}$ individuals are chosen at random from the population and given dose $d_i$ of compound B. Let $r_i^{\prime}$ denote the number of these individuals with response S.

We will use the following logistic regression models. Suppose that 
$r_1, \dots, r_m, r_1^{\prime}, \dots, r_m^{\prime}$ are independent. Also suppose that $r_i \sim \text{Binomial}(\text{n}_i, \text{p}_i)$ and $r_i^{\prime} \sim \text{Binomial}(\text{n}_i^{\prime}, \text{p}_i^{\prime})$ for $i = 1, \dots, m$.
Let $\text{logit}(\text{p}) = \log\big(\text{p} / (1 - \text{p})\big)$ for 
$0 < \text{p} < 1$.
Suppose that for any dose level $d$ of compound A, the probability $\text{p}$ of response S for a randomly chosen individual from the population is given by 
$\text{logit}(\text{p}) = \beta_1 + \beta_2 \, x$, 
where $x = \log_{10}(d)$. This implies that 
\begin{equation}
\label{ModelCompoundA}
\text{logit}(\text{p}_i) = \beta_1 + \beta_2 \, x_i \ \ \text{for} \ \ i=1, \dots, m.
\end{equation}
Also suppose that 
for any dose level $d^{\prime}$ of compound B, the probability $\text{p}^{\prime}$ of response S for a randomly chosen individual from the population is given by 
$\text{logit}(\text{p}^{\prime}) = \beta_3 + \beta_4 \, x^{\prime}$,
where $x^{\prime} = \log_{10}(d^{\prime})$. This implies that 
\begin{equation}
\label{ModelCompoundB}
\text{logit}(\text{p}_i^{\prime}) = \beta_3 + \beta_4 \, x_i \ \ \text{for} \ \ i=1, \dots. m,
\end{equation}
Let $\bm{y} = \big(r_1, \dots, r_m, r_1^{\prime}, \dots, r_m^{\prime}\big)$ and 
$\bm{\beta} = \big(\beta_1, \dots, \beta_4\big)$, so that this is a model of the 
type described in Section 3. 

Let $\text{ED}_{z}$ denote the log-dose $x$ of compound A for which the probability of response S for a randomly chosen individual from the population is $z/100$. Also let $\text{ED}_{z}^{\prime}$ denote the log-dose $x^{\prime}$ of compound B for which the probability of response S for a randomly chosen individual from the population is $z/100$. Suppose that the parameter of interest
$\theta = \text{ED}_{z} - \text{ED}_{z}^{\prime}$, for some given $z$ 
($0 < z < 100$).

Consider the case that for all possible dose levels
$d$ of compound A, the probability $p$ of response S for a randomly chosen individual from the population is the same as that for a dose level $d^{\prime} = \lambda d$ of compound B, for some fixed $\lambda > 0$.
Therefore, the log-dose $x = \log_{10}(d)$ of compound A leads to the same probability $p$ of response S for a randomly chosen individual from the population as the
log-dose $\log_{10}(\lambda d) = \log_{10}(\lambda) + \log_{10}(d)$ of compound B. Hence
$\text{logit}(p) = \beta_1 + \beta_2 \, x$ and 
\begin{equation*}
\text{logit}(p) = 
\text{logit}(p^{\prime}) 
= \beta_3 + \beta_4 \, \big(\log_{10}(\lambda) + \log_{10}(d) \big)
= \big(\beta_3 + \beta_4 \, \log_{10}(\lambda) \big) + \beta_4 \, x.
\end{equation*}
Therefore $\beta_2 = \beta_4$, so that the straight 
lines $\beta_1 + \beta_2 \, x$ and 
$\big(\beta_3 + \beta_4 \, \log_{10}(\lambda) \big) + \beta_4 \, x$
are parallel. 
This condition of ``parallelism'', i.e. that $\beta_2 = \beta_4$, greatly simplifies the statistical analysis.

We consider the case that, although the compounds A and B are thought 
\textsl{a priori} to be sufficiently similar that the hypothesis of ``parallelism'' is highly plausible, we are not certain that this hypothesis holds. In other words, suppose that we have uncertain prior information that the hypothesis of ``parallelism'' holds.

\section{Numerical illustration: quantal bioassay of Morphine and Amidone.}

We illustrate the properties of the confidence interval $\text{ACI}_{\text L}(\bm{y})$,
which utilizes the uncertain prior information that the hypothesis of ``parallelism''
holds, using data from \cite{Grewal1952}. This data was collected to compare the analgesic properties of Morphine and Amidone (also known as Methadone) in mice. 
A total of 616 homogeneous mice were randomly allocated to the groups shown in Table 1. In this table $x_i$ denotes $\log_{10}$ of the dose and $n_i$  and $n_i^{\prime}$ denote the number of mice give this log-dose of Morphine and Amidone, respectively. The experimenter recorded the number of shocks that could be applied to the tail of the mouse before it squeaked. If the number of shocks was four or more then the mouse was taken to give response S. In Table 1, $r_i$ 
denotes the number of mice (out of $n_i$ mice) with response S for the log-dose $x_i$ of Morphine. Similarly, $r_i^{\prime}$ denotes the number of mice (out of $n_i^{\prime}$ mice) with response S for the log-dose $x_i$ of Amidone.

\setcounter{table}{0}

\begin{table}[H]
	\caption{Quantal bioassay of Morphine and Amidone}
	\label{Grewal1952Data}
	\begin{tabular}{lllll}
		\hline
		\multicolumn{1}{c}{$\log_{10}$ dose} & \multicolumn{2}{c}{Morphine}                  & \multicolumn{2}{c}{Amidone}                   \\ \hline
		\multicolumn{1}{c}{$x_i$}    & \multicolumn{1}{c}{$n_i$} & \multicolumn{1}{c}{$r_i$} & \multicolumn{1}{c}{$n_i^{\prime}$} & \multicolumn{1}{c}{$r_i^{\prime}$} \\ \hline
		$0.18$ & $103$ & $19$ & $60$ & $14$ \\ \hline
		$0.48$ & $120$ & $53$ & $110$ & $54$ \\ \hline
		$0.78$ & $123$ & $83$ & $100$ & $81$ \\ \hline
	\end{tabular}
\end{table}

Suppose that the parameter of interest is 
$\theta = \text{ED}_{z} - \text{ED}_{z}^{\prime}$, where 
$\text{ED}_{z}$ and $\text{ED}_{z}^{\prime}$ are the log-doses of Morphine and Amidone, respectively, for which the probability of response S for a randomly chosen mouse is $z / 100$. Also suppose that our aim is to find a confidence interval for $\theta$ with minimum coverage probability 0.95.

Morphine and Amidone both belong to the family of drugs known as opioids. Opioids act on the brain in a particular way that can provide pain relief. Because Morphine and Amidone are both opioids, the hypothesis of ``parallelism'' is highly plausible. However, we are not certain that this hypothesis holds. In other words, we have uncertain prior information that the hypothesis of ``parallelism'' holds.

The models that we use for the Morphine and Amidone data are  
\eqref{ModelCompoundA} and \eqref{ModelCompoundB}, respectively, with $m = 3$, $\text{n}_1 = 103$,
$\text{n}_2 = 120$, $\text{n}_3 = 123$, $\text{n}_1^{\prime} = 60$,
$\text{n}_2^{\prime} = 110$ and $\text{n}_3^{\prime} = 100$. 
The parameter of interest is
\begin{align*}
\theta = g(\bm{\beta}) = \frac{1}{\beta_2}\left(\text{logit}\left( \frac{z}{100}\right) - \beta_1 \right) 
- \frac{1}{\beta_4}\left(\text{logit}\left( \frac{z}{100}\right) - \beta_3 \right).
\end{align*}
\textbf{Henceforth, we consider that case that $\boldsymbol{z = 60}$}.
Let 
$\tau = h(\bm{\beta}) = \beta_2 - \beta_4$.
The uncertain prior information is that $\tau = 0$.

\subsection{Local assessment of the coverage probability of a confidence interval}


We apply the procedure described in subsubsection \ref{Data_Based_Choice_Of_Beta_tilde} to the Morphine/Amidone data.
For this data, 
the maximum likelihood estimate  
$\widehat{\bm{\beta}} = (-2.0652, 3.6418, -2.0968, 4.4581)$.  
We find that 
$\widetilde{\bm{\beta}} 
= \big(\widetilde{\beta}_1, \dots , \widetilde{\beta}_p\big)$ is given  by 
\begin{equation*}
	\widetilde{\beta}_1 = \widehat{\beta}_1, \
	\widetilde{\beta}_2 = \frac{\widehat{\beta}_2 + \widehat{\beta}_4}{2}, \
	\widetilde{\beta}_3 = \widehat{\beta}_3, \
	\widetilde{\beta}_4 = \frac{\widehat{\beta}_2 + \widehat{\beta}_4}{2}.
\end{equation*}

The data for Morphine and Amidone come from independent experiments, so that the estimators $\big(\widehat{\beta}_1, \widehat{\beta}_2 \big)$ and $\big(\widehat{\beta}_3, \widehat{\beta}_4 \big)$
are independent. As a result of this, the inverse of the Fisher information matrix, $\Big(I(\bm{\beta}) \Big)^{-1}$, is block diagonal. 
Using the expression for the Fisher information matrix, in the context of a logistic regression model, given on 
page 116 of \cite{McCullaghNelder1989} we find that
%
\begin{equation*}
	\Big(I(\widetilde{\bm{\beta}}) \Big)^{-1} =
	\begin{bmatrix}
		\begin{array}{cc;{2pt/2pt}cc}
			0.086304 & -0.142229 & 0  & 0 \\ 
			-0.142229 & 0.280902 & 0 & 0    \\ \hdashline[2pt/2pt]
			0 & 0 &  0.132504 & -0.216338   \\ 
			0 & 0 & -0.216338 & 0.408074
		\end{array}
	\end{bmatrix}.
\end{equation*}
Therefore
\begin{align*}
	&\text{avar}\big(\widehat{\theta}; \widetilde{\bm{\beta}} \big)
	= \frac{\partial g(\widetilde{\bm{\beta}})}{\partial \bm{\beta}} \,
	\big(I(\widetilde{\bm{\beta}}) \big)^{-1}  
	\left(\frac{\partial g(\widetilde{\bm{\beta}})}{\partial \bm{\beta}} \right)^{\top}
	= 0.002333,
	\\
	&\text{avar}\big(\widehat{\tau}; \widetilde{\bm{\beta}} \big)
	= \frac{\partial h(\widetilde{\bm{\beta}})}{\partial \bm{\beta}} \,
	\big(I(\widetilde{\bm{\beta}}) \big)^{-1}  
	\left(\frac{\partial h(\widetilde{\bm{\beta}})}{\partial \bm{\beta}} \right)^{\top}
	= 0.688976,
	\\
	&\text{acov}\big(\widehat{\theta}, \widehat{\tau}; \widetilde{\bm{\beta}}\big)
	= \frac{\partial g(\widetilde{\bm{\beta}})}{\partial \bm{\beta}} \,
	\big(I(\widetilde{\bm{\beta}}) \big)^{-1}  
	\left(\frac{\partial h(\widetilde{\bm{\beta}})}{\partial \bm{\beta}} \right)^{\top}
	= -0.01603,
\end{align*}
so that $\rho (\widetilde{\bm{\beta}}) = -0.399855$.

We also find that 
\begin{equation*}
\bm{\beta}^* 
= \widetilde{\bm{\beta}} 
+ \frac{\big(\text{avar}(\widehat{\tau}; \widetilde{\bm{\beta}}) \big)^{1/2}}{2}
\big[0 \ \ 1 \ \ 0 \ -1\big]^{\top} \, \gamma_a.
\end{equation*}
Let $\tau^* = h(\bm{\beta}^*) = \beta_2^* - \beta_4^*$. Now let
$\gamma^* 
= \tau^* \big/ \big(\text{avar}(\widehat{\tau}; \widetilde{\bm{\beta}}) \big)^{1/2}$.
It follows that $\gamma^* = \gamma_a$, so that 
$\bm{\beta}^* = \big(\beta_1^*, \dots, \beta_p^*\big)$ is given by
\begin{equation}
\label{MorphineAmidoneLocalModel}
\beta_1^* =  \widehat{\beta}_1, \
\beta_2^* = \frac{\widehat{\beta}_2 + \widehat{\beta}_4}{2} + \frac{\tau^*}{2}, \
\beta_3^* =  \widehat{\beta}_3, \
\beta_4^* = \frac{\widehat{\beta}_2 + \widehat{\beta}_4}{2} - \frac{\tau^*}{2},
\end{equation}
where 
$\tau^* 
= \big(\text{avar}(\widehat{\tau}; \widetilde{\bm{\beta}}) \big)^{1/2}
\, \gamma^*$
 and $\gamma^* \in [-u, u]$. We deal with the choice of $u$ in the next subsection.


%

\subsection{Choice of $\boldsymbol{u}$}

For $\beta_4^* = 0$ it is impossible to determine $\text{ED}_z$ for any $0 < z < 100$. Furthermore, values of $\beta_4^* < 0$ seem impossible. Therefore, $\beta_4^* = 0$ is a boundary point for impossible values of $\beta_4^*$. Note that $\beta_4^* = 0$ when $\tau^* = \widehat{\beta}_2 + \widehat{\beta}_4$.
Similarly, for $\beta_2^* = 0$ it is impossible to determine $ED_z$ for any $0 < z < 100$. Furthermore, values of $\beta_2^* < 0$ seem impossible. Therefore, $\beta_2^* = 0$ is a boundary value for impossible values of $\beta_2^*$. Note that $\beta_2^* = 0$ when 
$\tau^* = - \big(\widehat{\beta}_2 + \widehat{\beta}_4 \big)$.
Therefore, $b$ must be less than $\widehat{\beta}_2 + \widehat{\beta}_4 = 8.0999$.

In fact, $u$ must be a good deal less than $\widehat{\beta}_2 + \widehat{\beta}_4$
for the profile likelihood confidence interval for $\theta^*$, with nominal coverage 0.95, not to have extremely large lengths for a substantial proportion of samples. This is evident from 
Table \ref{gamma_para3} which shows the values of $\tau^*$ and $\gamma^*$ and the percentage of simulation runs for which
the length of the profile likelihood confidence interval is greater than 1000 for $z = 60$ and $M=5000$ simulation runs. We have therefore chosen $u = 2.5$, so that we restrict attention to $\gamma^* \in \{-2.5, -2, \dots, 2, 2.5 \}$.
Note that $\gamma^* = 2.5$ corresponds to $\tau^* = 2.075$.
To get a sense of the difference in slopes that this allows, consider the following.
%
\begin{enumerate}

    \item[(i)] 
    
    Suppose that $\gamma^* = 0$, so that $\tau^* = 0$ and the 
    hypothesis of ``parallelism'' is satisfied. In this case, 
    $\beta_2^* = \beta_4^* =  4.0499$. Consequently,
    $\text{ED}_{60} = 0.6101$ and $\text{ED}_{60}^{\prime} = 0.6178$, so that 
    $\theta^* = -0.0078$.
    
        \item[(ii)] 
    
    Suppose that $\gamma^* = 2.5$, so that $\tau^* = 2.075$. In this case, $\beta_2^* = 5.0874$, $\beta_4^* = 3.0124$. Consequently, 
     $\text{ED}_{60} = 0.4856$ and $\text{ED}_{60}^{\prime} = 0.8306$, so that 
    $\theta^* = -0.345$.
    
     \item[(iii)] 
    
    Suppose that $\gamma^* = -2.5$, so that $\tau^* = -2.075$. In this case, $\beta_2^* = 3.0124$, $\beta_4^* = 5.0874$. Consequently, 
     $\text{ED}_{60} = 0.8202$ and $\text{ED}_{60}^{\prime} = 0.4918$, so that 
    $\theta^* = 0.3283$.
    
\end{enumerate}

%
\newcolumntype{g}{>{\columncolor{Gray}}c}
\begin{table}[H]
	\caption{The values of $\gamma^*$ and $\tau^*$ and the percentage  of simulation runs (row marked \%) for which
		the length of the profile likelihood confidence interval is greater than 1000 for $z = 60$ and $M=5000$ simulation runs.} 
	\label{gamma_para3}
	\setlength\tabcolsep{5pt} 
	\begin{tabular}{l|lllllggggg}
		\specialrule{.05em}{0em}{0em}
		$\gamma^*$ & $-5$ & $-4.5$ & $-4$ & $-3.5$ & $-3$ & $-2.5$ & $-2$ & $-1.5$ & $-1$ & $-0.5$ \\ 
		\specialrule{.05em}{0em}{0em} 
		$\tau^*$ & $-4.15$ & $-3.74$ & $-3.32$ & $-2.91$ & $-2.49$ & $-2.08$ & $-1.66$ & $-1.25$ & $-0.83$ & $-0.42$ \\ 
		\specialrule{.05em}{0em}{0em} 
		$\%$ & 2.10 & 1.08 & 0.16 & 0.02 & 0.01 & 0.00 & 0.00 & 0.00 & 0.00 & 0.00 \\ 
		\specialrule{.05em}{0em}{0em}
	\end{tabular}
	\setlength\tabcolsep{5pt} 
	\begin{tabular}{l|gggggglllll}
		\specialrule{.05em}{0em}{0em}
		$\gamma^*$ & 0 & 0.5 & 1 & 1.5 & 2 & 2.5 & 3 & 3.5 & 4 & 4.5 & 5 \\ 
		\specialrule{.05em}{0em}{0em} 
		$\tau^*$ & 0.00 & 0.42 & 0.83 & 1.25 & 1.66 & 2.08 & 2.49 & 2.91 & 3.32 & 3.74 & 4.15 \\ 
		\specialrule{.05em}{0em}{0em} 
		$\%$ & 0.00 & 0.00 & 0.00 & 0.00 & 0.00 & 0.00 & 0.08 & 0.53 & 1.50 & 3.92 & 7.51 \\ 
		\specialrule{.05em}{0em}{0em}
	\end{tabular}
\end{table}

\subsection{Monte Carlo simulation estimation of the \textsl{local coverage probabilities} and \textsl{scaled expected lengths} of $\boldsymbol{\text{I}_{\text{L}}(y^*; c)}$
and $\boldsymbol{\text{ACI}_{\text{L}}(y^*)}$} 

Suppose that $\bm{\beta}^*$ is given by 
\eqref{MorphineAmidoneLocalModel}, where $\gamma^*$
(and therefore $\tau^*$) is specified. 
Replace $\bm{\beta}$
by $\bm{\beta}^*$ in the models 
\eqref{ModelCompoundA} and \eqref{ModelCompoundB}
that we use for the Morphine and Amidone data, respectively. For these models, 
 $k = 3$, $n_1 = 103$,
$n_2 = 120$, $n_3 = 123$, $n_1^{\prime} = 60$,
$n_2^{\prime} = 110$ and $n_3^{\prime} = 100$. 
Let $\bm{y}^*$ denote the response vector.

Our Monte Carlo simulation results show that 
the local coverage and scaled expected length properties of  
$\text{ACI}_{\text{L}}(\bm{y}^*)$ are superior to these properties 
for $\text{ACI}_{\text{W}}(\widehat{\bm{\beta}}^*)$. 
Consequently, the description of these properties for
$\text{ACI}_{\text{W}}(\widehat{\bm{\beta}}^*)$
have been relegated to Section S2 of the Supplementary Material.
For the remainder of the paper, we deal only with these properties for $\text{ACI}_{\text{L}}(\bm{y}^*)$.

\subsubsection{Monte Carlo simulation estimation of the 
\textsl{local coverage probabilities}}

The Monte Carlo simulation estimation of the \textsl{local coverage probabilities} of $\text{I}_{\text{L}}(\bm{y}^*; c)$, for given $c$,
and $\text{ACI}_{\text{L}}(\bm{y}^*)$ are very similar. 
Let $\theta^* = g(\bm{\beta}^*)$. We carry out $M$ independent simulation runs.
The $k$th simulation run generates an observation of $\bm{y}^*$.
We make Assumption A (stated in Appendix A) with $\bm{\beta}$ and $\bm{y}$
replaced by $\bm{\beta}^*$ and $\bm{y}^*$, respectively.
The Monte Carlo simulation results reported in Section S5 of the Supplementary Material provide evidence in favour of the correctness of this assumption.

We estimate the coverage probability 
$P_{\bm{\beta}^*}\big(\theta^* \in 
\text{I}_{\text{L}}(\bm{y}^*; c) \big)$, for given $c$, as follows.
On the $k$th simulation run we record
$\bm{1}\big(-z_{1 - \alpha/2} \leq r_1(\theta^* \, | \, \bm{y}^*) \leq z_{1 - \alpha/2}\big)$. Using the recorded results for the $M$ simulation runs, 
we estimate this coverage probability and the standard error of this estimate in the obvious way.

We estimate the coverage probability 
$P_{\bm{\beta}^*}\big(\theta^* \in 
\text{ACI}_{\text{L}}(\bm{y}^*) \big)$ as follows.
On the $k$th simulation run we record
\begin{equation}
\label{eq:Indicator_Cov_Prob_ACI_L}
 \bm{1}   \Big(b_{\rho(\widehat{\bm{\beta}}^*)} \big(r_2(\bm{y}^*) \big) 
- s_{\rho(\widehat{\bm{\beta}}^*)}\big(r_2(\bm{y}^*)\big) \leq 
r_1(\theta^* \, | \, \bm{y}^*) \leq 
b_{\rho(\widehat{\bm{\beta}}^*)} \big(r_2(\bm{y}^*)\big) 
+ s_{\rho(\widehat{\bm{\beta}}^*)} \big(r_2(\bm{y}^*)\big)
\Big).
\end{equation}
Using the recorded results for the $M$ simulation runs, 
we estimate this coverage probability and the standard error of this estimate in the obvious way.

\begin{figure}[H]
	\centering
	\includegraphics[scale=0.8]{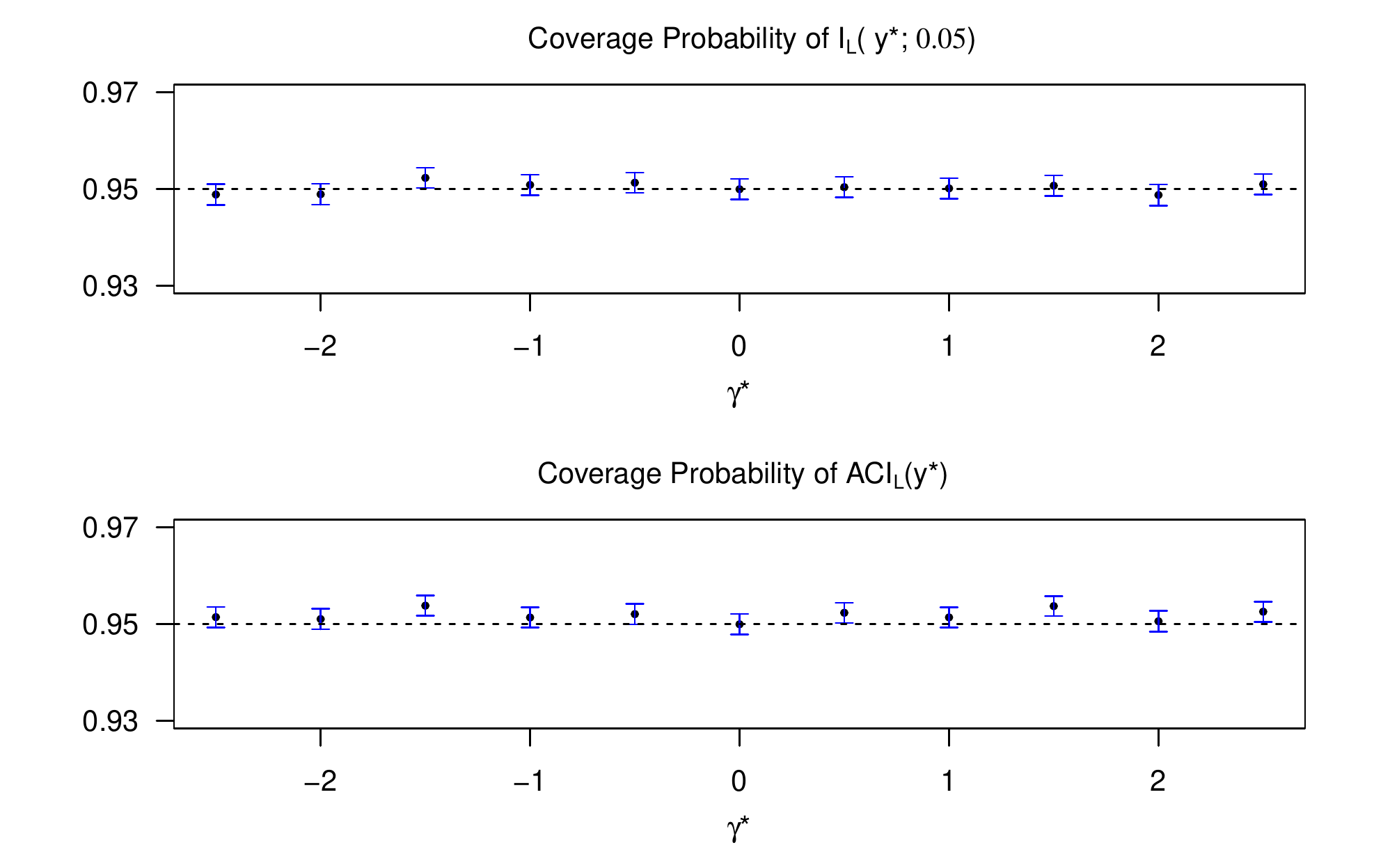} 
	\vspace{-1cm}
	\caption{Approximate 95\% confidence intervals for the coverage probabilities of the confidence intervals $\text{I}_{\text{L}}(\bm{y}^*; 0.05)$ (top panel) and $\text{ACI}_{\text{L}}(\bm{y}^*)$ (bottom panel), both with nominal coverage $0.95$, for $\gamma^* \in \{-2.5, -2, \dots, 2, 2.5 \}$.}
	\label{CovProb_IL_ACI_L}
\end{figure}

The top panel of Figure \ref{CovProb_IL_ACI_L} presents approximate 95\% confidence intervals for the coverage probability of the confidence interval 
$\text{I}_{\text{L}}(\bm{y}^*; 0.05)$,
which has nominal coverage 0.95, evaluated at 
$\gamma^* \in \{-2.5, -2, \dots, 2, 2.5 \}$.
The bottom panel of this figure presents approximate 95\% confidence intervals for the coverage probability of the confidence interval $\text{ACI}_{\text{L}}(\bm{y}^*)$, which has nominal coverage
0.95, evaluated on the same set of values of $\gamma^*$.
For both of these panels,  the number of simulation runs $M = 40,000$. 
These figures show that both 
$\text{I}_{\text{L}}(\bm{y}^*; 0.05)$
and 
$\text{ACI}_{\text{L}}(\bm{y}^*)$
have good local coverage properties.

\subsubsection{Monte Carlo simulation estimation of the \textsl{local scaled expected length} of $\boldsymbol{\text{ACI}_{\text{L}}(y^*)}$} 

Let
\begin{equation*}
q_{\text{L}}^* =  
\frac{\text{length of} \ \text{ACI}_{\text L}\big(\bm{y}^*\big) }
{\text{length of} \ \text{I}_{\text{L}}(\bm{y}^*; \widetilde{c}) \ \text{computed from the same data}},
\end{equation*}
where $\widetilde{c}$ is such that the \textsl{local minimum coverage probabilities} of $\text{ACI}_{\text L}\big(\bm{y}^*\big)$ and $\text{I}_{\text{L}}(\bm{y}^*;\widetilde{c})$
are the same.
The \textsl{local scaled expected length} of 
$\text{ACI}_{\text W}\big(\widehat{\bm{\beta}}^*\big)$ was defined in subsection 3.4.
The \textsl{local scaled expected length} of $\text{ACI}_{\text L}\big(\bm{y}^*\big)$
is similarly defined to be $E_{\bm{\beta}^*}(q_{\text{L}}^*)$.

We computed $\widetilde{c}$ using the method described in Section B.1 of Appendix B, with $M^{\prime} = 10,000$. We then used Monte Carlo simulation to estimate $E_{\bm{\beta}^*}(q_{\text{L}}^*)$ as follows.
We carry out $M$ independent simulation runs. On the $k$th simulation run we generate an observation $q_{\text{L}}^*(k)$ of $q_{\text{L}}^*$.
We estimate $E_{\bm{\beta}^*}(q_{\text{L}}^*)$ by 
$\sum_{k=1}^M q_{\text{L}}^*(k) \big / M$.
The analysis of the distribution of $q_{\text{L}}^*$, given in Section S4 of the Supplementary Material, shows that 
this distribution does not have any long or heavy tails.

\begin{figure}[H]
	\centering
	\includegraphics[scale=0.75]{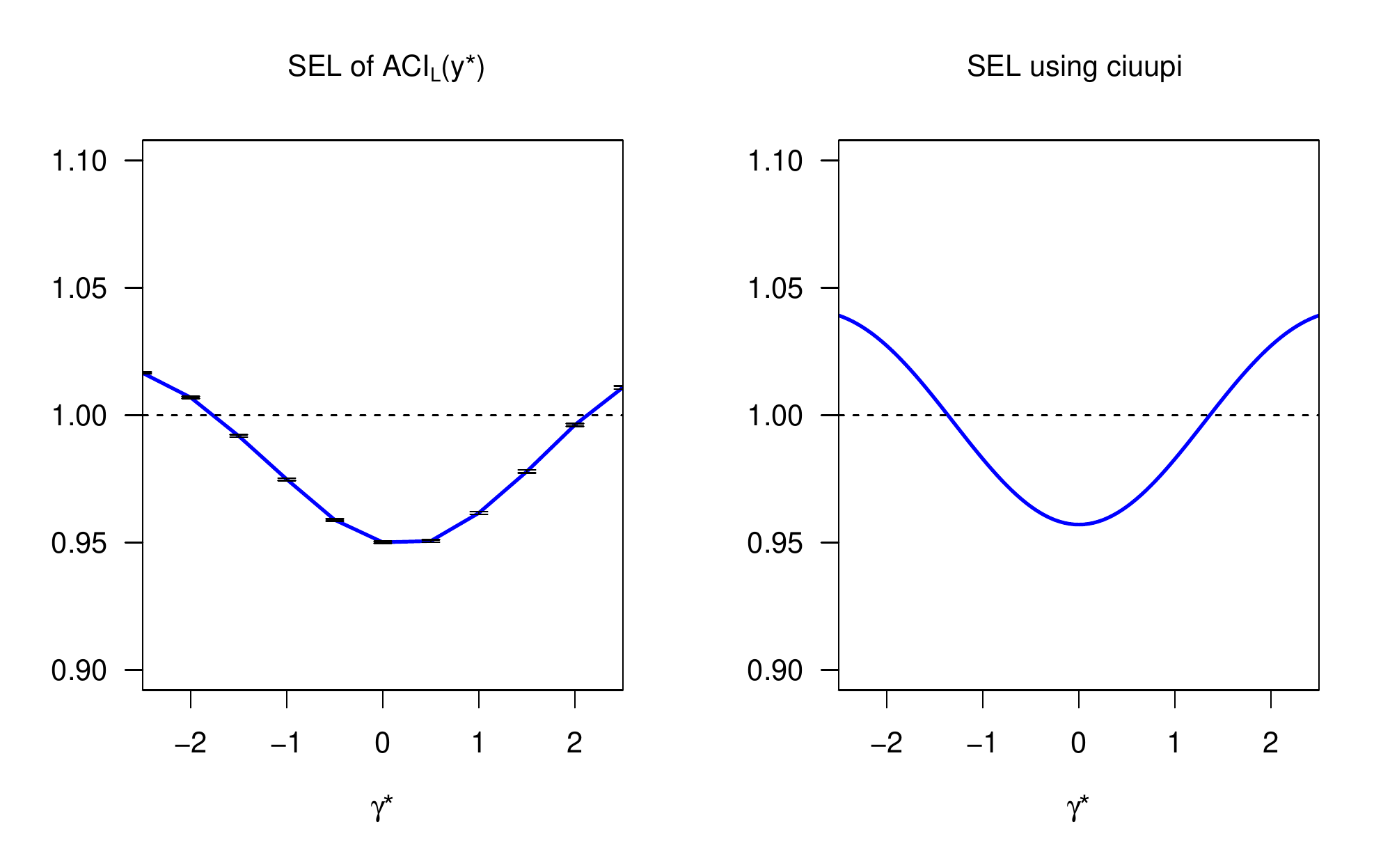}
	\vspace{-0.5cm}
	\caption{The left panel 
presents approximate 95\% confidence intervals for the \textsl{local scaled expected length} of the confidence interval $\text{ACI}_{\text L}\big(\bm{y}^*\big)$,
with nominal coverage 0.95, evaluated at $\gamma^* \in \{-2.5, -2, \dots, 2, 2.5 \}$. 
 The right panel is the graph of $SEL(\gamma^*; \rho(\widetilde{\bm{\beta}}))$ for the confidence interval
	$\text{CI}(b_{\rho(\widetilde{\bm{\beta}})}, s_{\rho(\widetilde{\bm{\beta}})})$,  found using 
	the \texttt{R} package \texttt{ciuupi}.}
	\label{SEL_mean_ACIvsWald}
\end{figure}

The left panel of Figure \ref{SEL_mean_ACIvsWald} 
presents approximate 95\% confidence intervals for the \textsl{local scaled expected length} of the confidence interval $\text{ACI}_{\text L}\big(\bm{y}^*\big)$,
with nominal coverage 0.95, evaluated at each
$\gamma^* \in \{-2.5, -2, \dots, 2, 2.5 \}$ using $M = 40,000$
simulation runs. 
These approximate 95\% confidence intervals were found using the simplifying 
approximation that $\widetilde{c}$ is computed without error. 
This panel shows that the confidence interval 
$\boldsymbol{\text{ACI}_{\text{L}}(y^*)}$
utilizes the uncertain prior information that $\tau^* = 0$.

\subsubsection{Large sample approximation to the \textsl{local coverage probability} and  scaled expected length of the confidence interval $\text{ACI}_{\text{L}}(y^*)$ } 
\label{AnalysisFromciuupi}

\begin{figure}[H]
	\centering
	\includegraphics[scale=0.65]{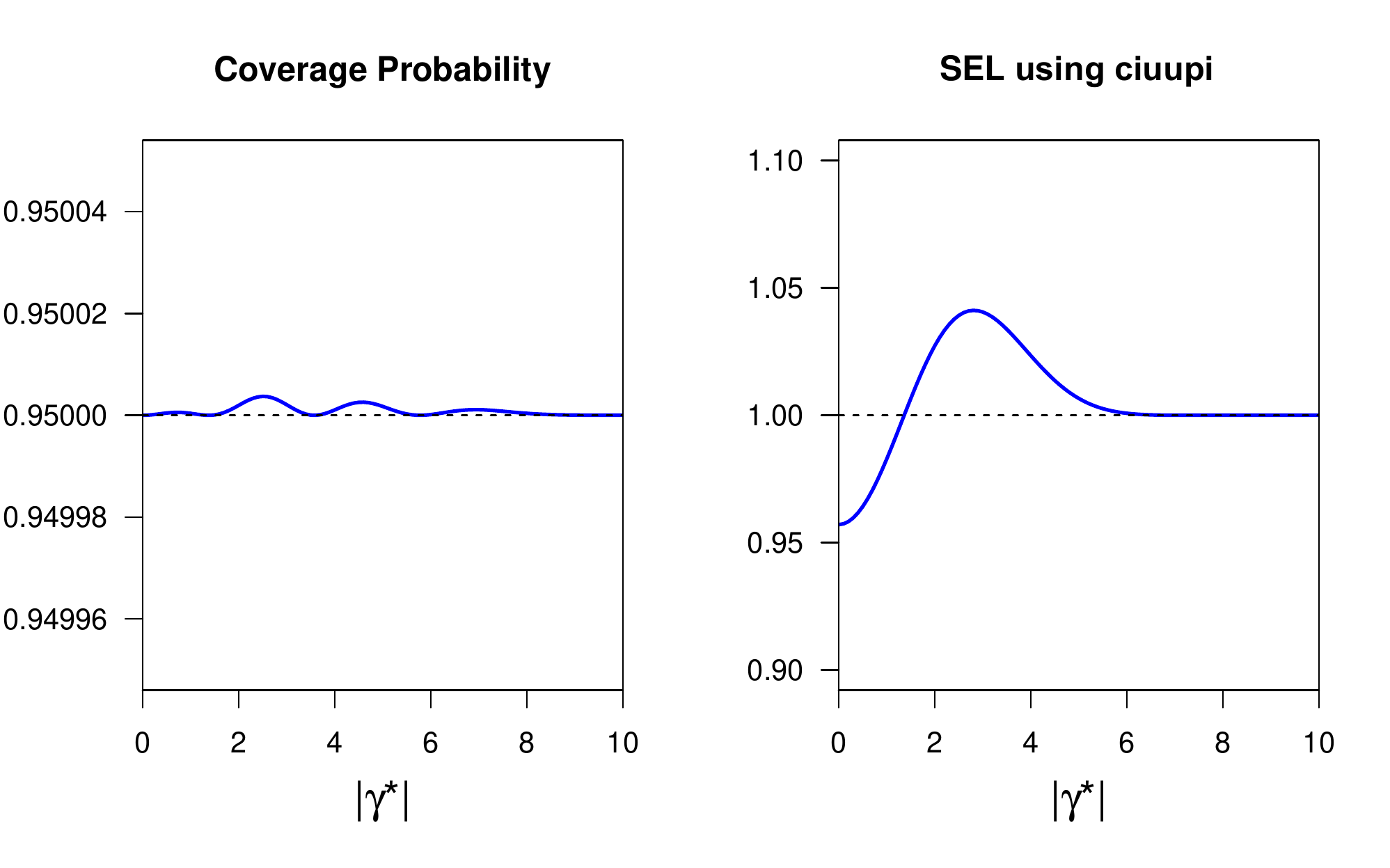}
	\vspace{-0.8cm}
	\caption{Graphs of the coverage probability $CP(\gamma^*; \rho(\widetilde{\bm{\beta}}))$ and the scaled expected length $SEL(\gamma^*; \rho(\widetilde{\bm{\beta}}))$ for the confidence interval
	$\text{CI} \big(b_{\rho(\widetilde{\bm{\beta}})}, s_{\rho(\widetilde{\bm{\beta}})} \big)$, found using 
	the \texttt{R} package \texttt{ciuupi}.}
	\label{Fig:CovProb_SEL}
\end{figure}

Application of the large sample approximations given in 
subsubsection 
\ref{Data_Based_Choice_Of_Beta_tilde}
and subsection
\ref{Sect:DefnLocalScaledExpectedLength} to the Morphine/Amidone data gives the following results. 
The coverage probability and the scaled expected length of 
the confidence interval $\text{ACI}_{\text{L}}(y^*)$ 
are approximated
by the coverage probability $CP(\gamma^*; \rho(\widetilde{\bm{\beta}}))$ and the scaled expected length 
$SEL(\gamma^*; \rho(\widetilde{\bm{\beta}}))$ of the confidence interval $\text{CI}(b_{\rho(\widetilde{\bm{\beta}})}, s_{\rho(\widetilde{\bm{\beta}})})$ computed using the   \texttt{R} package \texttt{ciuupi}. Recall that 
$\rho(\widetilde{\bm{\beta}}) = -0.399855$
for the Morphine/Amidone data, 

Suppose that $1 - \alpha = 0.95$. Graphs of the coverage probability $CP(\gamma^*; \rho(\widetilde{\bm{\beta}}))$ and the scaled expected length $SEL(\gamma^*; \rho(\widetilde{\bm{\beta}}))$, considered as functions of $|\gamma^*| \in [0, 10]$, are shown in  
Figure \ref{Fig:CovProb_SEL}.
The right panel of Figure \ref{SEL_mean_ACIvsWald}
is a graph of $SEL(\gamma^*; \rho(\widetilde{\bm{\beta}}))$, considered as a function of $\gamma^* \in \{-2.5, -2, \dots, 2, 2.5 \}$. The left and right hand panels of Figure \ref{SEL_mean_ACIvsWald} show very similar qualitative features.

\section{Discussion}

For both designed experiments and observational studies, it is widely believed that the higher the order of an interaction term in the model the more likely it is that this term differs negligibly from zero.
Consider the case that (a) the responses are either binomial, negative binomial or Poisson distributed, (b) the parameter of interest is a scalar and (c) there is a single highest order interaction term. If we have a very strong belief that this interaction term differs negligibly from zero then we may simply choose to omit this term from the model. However, there will be circumstances in which we believe that this term differs negligibly from zero, while being uncertain of this. In these circumstances, a confidence interval that utilizes this uncertain prior information, constructed using the method we have described, is an attractive option.

For the last few decades a very active and practically-important area of research has been on frequentist confidence regions that include some aspect of model selection or model weighting. The connections between this research and the research on frequentist confidence regions that utilize uncertain prior information run deep. As astutely pointed out by the econometrician
\citeauthor{Leamer1978} (\citeyear{Leamer1978}, chapter 5), preliminary data-based model selection may be motivated by a desire to utilize uncertain prior information in subsequent inference. However, confidence intervals constructed after data-based model selection typically have very undesirable properties, see \cite{Kabaila2009_ISR} for a review. They certainly fail to utilize the uncertain prior information that may motivate their use. \cite{Leamer1978} tries to elicit the uncertain prior information that underlies a number of preliminary data-based model selection procedures. He then adopts a Bayesian approach to the proper incorporation of this prior information into subsequent inference. The work of Kabaila and co-authors
listed in the references may be considered, at least to some extent, to be a continuation of the program instituted by Leamer, but using frequentist methods for the inference of interest being a confidence region.
Another connection between post-model-selection confidence intervals and confidence intervals that utilize uncertain prior information is that the former have been modified to construct the latter, see e.g. 
\cite{Cohen1972} and \cite{KabailaGiri2009JSPI}.

We believe that Leamer's observation also applies to the frequentist model averaged confidence intervals described by \cite{BucklandEtAl1997}, \cite{FletcherTurek2011}
and \cite{TurekFletcher2012}
and the confidence intervals centered on a bootstrap smoothed (or bagged; \citeauthor{Breiman1996}, \citeyear{Breiman1996})
post-model-selection estimator described by \cite{Efron2014}.
In other words, these confidence intervals appear to be motivated by a desire to utilize uncertain prior information. The success of these confidence intervals in achieving this aim has been mixed, see \cite{KabailaWelshAbeysekera2016}, \cite{KabailaWelshMainzer2017}, \cite{Kabaila2018},
\cite{KabailaWelshWijethunga2020_JSPI}, \cite{KabailaWijethunga2019ANZJS} and
 \cite{KabailaWijethunga2019Stat}.



\appendix

\section{Appendix: Expressions for the coverage probabilities of $\boldsymbol{\text{I}_{\text{L}}(y; c)}$ and $\boldsymbol{\text{ACI}_{\text L}(y)}$ 
that do not require the computation of the endpoints of these confidence intervals}

We make the following assumption.

\smallskip

\noindent Assumption A: \
For the chosen true value of $\bm{\beta}$, 
there is a set ${\cal Y}_{\bm{\beta}}$ of values of $\bm{y}$ such that (a)
the probability that $\bm{y} \in {\cal Y}_{\bm{\beta}}$ is very close to 1 and 
(b) $r_1(\theta^{\prime}\, | \, \bm{y})$ is a decreasing function of $\theta^{\prime}$ for all $\bm{y} \in {\cal Y}_{\bm{\beta}}$. 

The coverage probability of 
$\text{I}_{\text{L}}(\bm{y}; c)$
can be computed to a very good approximation, without computing the endpoints of this confidence interval, as follows.
Since
\begin{equation*}
\left\{\widehat{\theta}_l \leq \theta \leq \widehat{\theta}_u \right\} 
= \left\{-z_{1 - \alpha/2} \leq r_1(\theta \, | \, \bm{y}) \leq z_{1 - \alpha/2}\right\},
\end{equation*}
the coverage probability of 
$\text{I}_{\text{L}}(\bm{y}; c)$ is, to a very good approximation, given by 
$P \big(-z_{1 - \alpha/2} \leq r_1(\theta \, | \, \bm{y}) \leq z_{1 - \alpha/2}\big)$.

Similarly, the coverage probability of $\text{ACI}_{\text L}(\bm{y})$ can be computed
to a very good approximation, without computing the endpoints of this confidence interval, as follows.
Since
\begin{equation*}
\left\{\widetilde{\theta}_l \leq \theta \leq \widetilde{\theta}_u\right\} 
= \Big\{b_{\rho(\widehat{\bm{\beta}})} \big(r_2(\bm{y}) \big) 
- s_{\rho(\widehat{\bm{\beta}})}\big(r_2(\bm{y})\big) \leq 
r_1(\theta \, | \, \bm{y}) \leq 
b_{\rho(\widehat{\bm{\beta}})} \big(r_2(\bm{y})\big) 
+ s_{\rho(\widehat{\bm{\beta}})} \big(r_2(\bm{y})\big)
\Big\},
\end{equation*}
the coverage probability of $\text{ACI}_{\text L}(\bm{y})$ is, to a very good
approximation, given by 
\begin{equation*}
P \Big(b_{\rho(\widehat{\bm{\beta}})} \big(r_2(\bm{y}) \big) 
- s_{\rho(\widehat{\bm{\beta}})} \big(r_2(\bm{y})\big) \leq r_1(\theta \, | \, \bm{y}) \leq 
b_{\rho(\widehat{\bm{\beta}})} \big(r_2(\bm{y})\big) 
+ s_{\rho(\widehat{\bm{\beta}})} \big(r_2(\bm{y})\big)\Big).
\end{equation*}

\section{Appendix: Computation of $\widetilde{c}$}

The method used to compute $\widetilde{c}$ such that the \textsl{local minimum coverage
probabilities} of $\text{ACI}_{\text{L}}(\bm{y}^*)$ 
and 
$\text{I}_{\text{L}}(\bm{y}^*; \widetilde{c})$ are the same is very similar
to the 
method used to compute $\widetilde{c}$ such that the \textsl{local minimum coverage
probabilities} of 
 $\text{ACI}_{\text W}\big(\widehat{\bm{\beta}}^*\big)$ and $\text{I}_{\text{W}}(\bm{y}^*;\widetilde{c})$
are the same.
For the sake of brevity, we describe only the latter method.

\subsection{Monte Carlo simulation estimation of the \textsl{minimum coverage probabilities} of $\text{ACI}_{\text{W}}(\widehat{\boldsymbol{\beta}}^*)$
and $\text{I}_{\text{W}}(\boldsymbol{y}^*;c)$}
\label{sect:MC_Est_Min_CP_ACI_W}

The computation of the \textsl{local minimum coverage probabilities} of $\text{ACI}_{\text{W}}(\widehat{\boldsymbol{\beta}}^*)$
and $\text{I}_{\text{W}}(\bm{y}^*;c)$, for given $c$, are very similar. For the sake of brevity, we describe only the latter.

If we estimate the \textsl{local minimum coverage probability} of $\text{I}_{\text{W}}(\bm{y}^*;c)$ by choosing the smallest of the estimated coverage probabilities,
for $\gamma^* \in \{-u, -u + \delta, \dots, u - \delta, u \}$, then this estimate will be biased downwards. We therefore use the following three step process.

\medskip

\noindent \underline{Step 1} : Estimate the coverage probability 
for each $\gamma^* \in \{ -u, -u + \delta, \dots, u - \delta, u \}$,
where $\delta = u / 5$. We use $M^{\prime}$ simulation runs for each value of $\gamma^*$. Pick the 3 values of $\gamma^*$ that have the smallest estimated coverage probability.

\noindent \underline{Step 2} : For the 3 values of $\gamma^*$ chosen in Step 1, run new simulations using $10M^{\prime}$ simulation runs for each value of $\gamma^*$. Choose the value of $\gamma^*$ (out of these 3 values) that minimizes the estimated coverage probability.

\noindent \underline{Step 3} : For the value of $\gamma^*$ chosen in Step 2, run a new simulation with $100M^{\prime}$ simulation runs to obtain the final estimate of the \textsl{local minimum coverage probability}.

\noindent This technique is a variant of the technique used in Section 3.1 of \cite{KabailaLeeb2006}.

\subsection{Use of a fitted straight line in $c$ to compute $\widetilde{c}$}

Note that 
\begin{equation*}
  P_{\bm{\beta}^*} \big( \theta^* \in \text{I}_{\text{W}}(\bm{y}^*;c) \big)  
  = P_{\bm{\beta}^*} \left( - z_{1 - c/2} \le 
  \frac{\widehat{\theta}^* - \theta^*}{\big(  \text{avar}(\widehat{\theta};
  \widehat{\bm{\beta}}^*\big)^{1/2} }
  \le z_{1 - c/2}
  \right)
  \approx 1 - c.
\end{equation*}
Consequently, 
to compute $\widetilde{c}$, we make the very reasonable assumption
that the \textsl{local minimum coverage probability} of $\text{I}_{\text{W}}(\bm{y}^*;c)$ is approximately a straight line function of $c$,
for $c$ close to $\alpha$. We choose 3 values of $c$: $\alpha - \delta_c, \alpha$ and $\alpha + \delta_c$, where $\delta_c$ is a judiciously-chosen small positive number. We then fit a straight line to 
 the \textsl{local minimum coverage probability} of $\text{I}_{\text{W}}(\bm{y}^*;c)$ 
 evaluated at these 3 values of $c$ to compute $\widetilde{c}$.

\end{document}